\documentclass[useAMS,usenatbib]{mn2e}
\usepackage{graphicx}
\usepackage[normalem]{ulem}
\usepackage{xcolor}
\usepackage[]{inputenc,amssymb}

\def \be{\begin{equation}}
\def \ee{\end{equation}}
\newcommand       \ba           {\begin{eqnarray}}
\newcommand       \ea           {\end{eqnarray}}
\def \bea{\begin{eqnarray}}
\def \eea{\end{eqnarray}}

\newcommand{\comments}[1]{}

\definecolor{webgreen}{rgb}{0,.5,0}
\definecolor{webbrown}{rgb}{.6,0,0}
\usepackage[pdfpagelabels]{hyperref}
\hypersetup{%
   colorlinks=true,hyperfootnotes=false,%
   breaklinks=true,%
   plainpages=false, bookmarksnumbered, bookmarksopen=true,
   bookmarksopenlevel=1,%
   urlcolor=webbrown, linkcolor=webbrown, citecolor=webgreen,
   }

\title[]{AR Sco as a possible seed of highly magnetised white dwarf}   

\author[B. Mukhopadhyay, A. R. Rao, T. S. Bhatia]
{Banibrata Mukhopadhyay,$^{1}$
A. R. Rao,$^{2}$
Tanayveer Singh Bhatia$^1$
\\
$^{1}$Department of Physics, Indian Institute of Science, Bangalore 560012; bm@iisc.ac.in ,
tanayveer1@gmail.com\\
$^2$Department of Astrophysics and Astronomy, Tata Institute of Fundamental Research,
Mumbai 40005; arrao@tifr.res.in
}

\begin{document}
\maketitle

\label{firstpage}

\begin{abstract}
We explore the possibility that the recently discovered white dwarf pulsar AR Sco 
acquired its high spin and magnetic field due to repeated episodes of accretion 
and spin-down. An accreting white dwarf   can lead to 
 a larger mass and consequently a  smaller radius thus causing  
an enhanced rotation period and magnetic field. This spinning magnetic white dwarf
temporarily can inhibit accretion, spin down, and, eventually, the accretion can start
again due to the shrinking of the binary period by  gravitational radiation. A 
repeat of the above cycle can eventually lead to a high magnetic field white dwarf, recently
postulated to be the reason for over-luminous type Ia supernovae. We also
point out that these high magnetic field
spinning white dwarfs are attractive sites for gravitational radiation.
\end{abstract}


\section{Introduction}           
\label{sect:intro}

Recently, we had shown that 
 the mass-radius ($M-R$) relation of Chandrasekhar has to be suitably
modified in the case of  
 highly magnetised white dwarfs, namely B-WDs, leading to significantly higher
mass limit for white dwarfs. By means of both analytical calculations and
numerical modelling, we carried out a systematic study of how a strong magnetic 
field affects the structure and properties of the underlying white dwarf in 
a variety of ways. We progressed from
constructing simplistic to more rigorous and self-consistent models 
(see, e.g., \citealt{prl,dmr,sathya,bmrao}). After our initiation, other
independent groups also 
examined the implications of high magnetic fields for the mass-radius 
relation of white dwarfs 
(e.g. \citealt{liu,bb15,belya,schram}).

 The prime motivation of the B-WD model was to  explain the  peculiar over-luminous type 
Ia supernovae (\citealt{dm15}). The question, however, remained about the
mechanism by which white dwarfs can attain high magnetic fields. The related
question is to explore the observational consequences of the existence of  a large
number of B-WDs and proto-B-WDs. Since an increase in the mass of white dwarf leads to a
decrease in its radius, the mechanism of accretion leading to an increase in its magnetic
field by flux freezing, and hence leading to B-WDs, was also qualitatively explored to explain some other observed
phenomena. They are 
  soft gamma-ray repeaters (SGRs) and anomalous X-ray pulsars (AXPs), 
particularly the ones exhibiting 
high X-ray luminosities (\citealt{bmrao}) which posit problem on the neutron star based model (see, e.g., \citealt{mereghetti}), and 
some white dwarf pulsars e.g. GCRT J1745-3009 (\citealt{bing,bmrao}).

The recent discovery of 
a rotating magnetised white dwarf
(\citealt{marshnature}) in AR Scorpii (AR Sco) demonstrates that some  white dwarfs do 
acquire high magnetic field and high spin period. The evolutionary scenario of
AR Sco, though not fully explored (see, however, \citealt{bes}) in the literature,
could involve 
the scenario envisaged for the generation of high magnetic field in B-WDs
(accretion resulting in smaller radius). In this paper, we explore this possibility and
point out that AR Sco appears to be a very suitable candidate to be a seed B-WD.
  
In the next section, we discuss the $M-R$ relations of moderately magnetised B-WDs, with 
surface (and central) magnetic fields
similar to those inferred for AR Sco.
Subsequently, in \S 3 we will discuss the possible time evolution of AR Sco and how 
being in a $M-R$ trajectory of moderately magnetised white 
dwarfs, it is expected to switch to a $M-R$ trajectory of B-WDs. We will also comment on
the possible emission and detection of gravitational waves from it and in general B-WDs in \S 4, apart
from the fact that the glitches and outbursts seen in SGRs/AXPs can be explained in the B-WD
premises. Finally we end with a conclusion in \S 5.


\section{Moderately magnetized white dwarfs: pulsar AR Sco}
\label{massradius}

AR Sco has been shown to be a rotating magnetised white dwarf
(\citealt{marshnature}). This is argued based on various properties of it, e.g. increasing optical flux
detected in radio, higher spin-down power compared to electromagnetic radiation, no obvious sign 
of accretion, broadband spectrum characteristic of synchrotron radiation requiring relativistic 
electrons etc. The underlying white dwarf/cool star binary system emits radiation from X-ray to radio,
which is pulsating in brightness on a period $P=1.97$ min and period derivative 
${\dot P}\sim 4\times 10^{-13}$ sec sec$^{-1}$. The maximum and mean luminosities of 
AR Sco are $\sim 6.3\times10^{32}$ erg~s$^{-1}$ and $\sim 1.7\times 10^{32}$ erg~s$^{-1}$ respectively. The mass
$M$ of the white dwarf inferred to be in the range $0.8-1.29 M_\odot$, where $M_\odot$ is the mass of Sun.  
For the corresponding radius $R=7000-3200$ km, the spin-down power turns out to be $L_{\dot{\nu}}\sim 1.5\times 
10^{32-33}$ erg~s$^{-1}$,
which is adequate to explain its luminosity mentioned above. However, a neutron star based model with
its typical mass and radius, $L_{\dot{\nu}}$ turns out to be much smaller $\sim 10^{28}$ erg~s$^{-1}$,
which rules it out to be a spinning down neutron star. In the framework of an accreting 
compact object (which indeed does not have any sign in it), a neutron star requires 
the accretion rate $\dot{M}\sim 10^{-14} M_\odot$Yr$^{-1}$, while a white dwarf needs to have a much 
higher rate $\dot{M}\sim 1.3\times 10^{-11} M_\odot$Yr$^{-1}$ which is very high to see the Doppler-broadband
emission lines from accretion, but AR Sco shows features from M-star.

Keeping all the features in mind, AR Sco is confirmed to be a white dwarf rather than a neutron star,
which qualifies it to be the first ever detected white dwarf radio pulsar. 
Its surface magnetic field can be estimated by assuming the standard dipole
radiation from rotating magnetised compact objects (e.g. \citealt{bmrao}, see \S 3
for details) as 
$B_s\sim \sqrt{5c^3MP\dot{P}/4\pi^2R^4\sin^2\alpha}\sim 6\times 10^{8-9}$ G
with the angle between the spin and magnetic axes $\alpha=90$ degree, where $c$ is 
the speed of light. However, for smaller $\alpha$,
which cannot be ruled out (see, e.g., \citealt{tongxu12}), $B_s$ could exceed $10^{11}$ G. 
As a range of mass is inferred for AR~Sco from observation which corresponds to
a range of radius, a range of $B_s$ is inferred. Hence, 
central field $B_c$ could even be $10^{14}$ G. Such $B_s$ and $B_c$ could influence 
stellar structure due to the effect of magnetic pressure, however are small enough to practically affect
electron-degenerate matter. Hence, Chandrasekhar's equation of state (\citealt{chandra35}) would suffice.

In the presence of stronger magnetic fields, even for $B_c\sim 10^{14}$ G as is expected in AR Sco,
white dwarfs' structure and $M-R$ relation may deviate from that of Chandrasekhar's theory.
In fact, such white dwarfs (and B-WDs) need not necessarily be spherical, depending on the field
values. Such B-WDs were explored in past by us and other groups (\citealt{dm15,schram,bb15}), when
the focus was mainly to investigate how massive B-WDs could be in the presence of various kinds
of fields. For the present purpose, when field magnitudes are rather restricted from observational
inference, we construct $M-R$ relations of B-WDs based on publicly available 
LORENE code\footnote{http://www.lorene.obspm.fr} 
(\citealt{lorene1,lorene2}).
By construction, LORENE describes compact stars, in the general relativistic framework,
having purely poloidal fields. Such B-WDs could
be oblate, depending on field strength, with a smaller equatorial radius ($R_e$) compared to their weakly
magnetic counterparts (see, e.g., \citealt{sathya}), unlike toroidally dominated 
B-WDs which are prolate spheroids. Their mass is also restricted to $\sim 2M_\odot$, unlike 
their toroidally dominated counterparts (\citealt{dm15,sathya}). However, such massive B-WDs 
are possible for $B_c\gtrsim 3\times 10^{14}$ G. As mentioned above, fields in AR Sco are smaller.

\begin{figure}
        \includegraphics[scale=0.35]{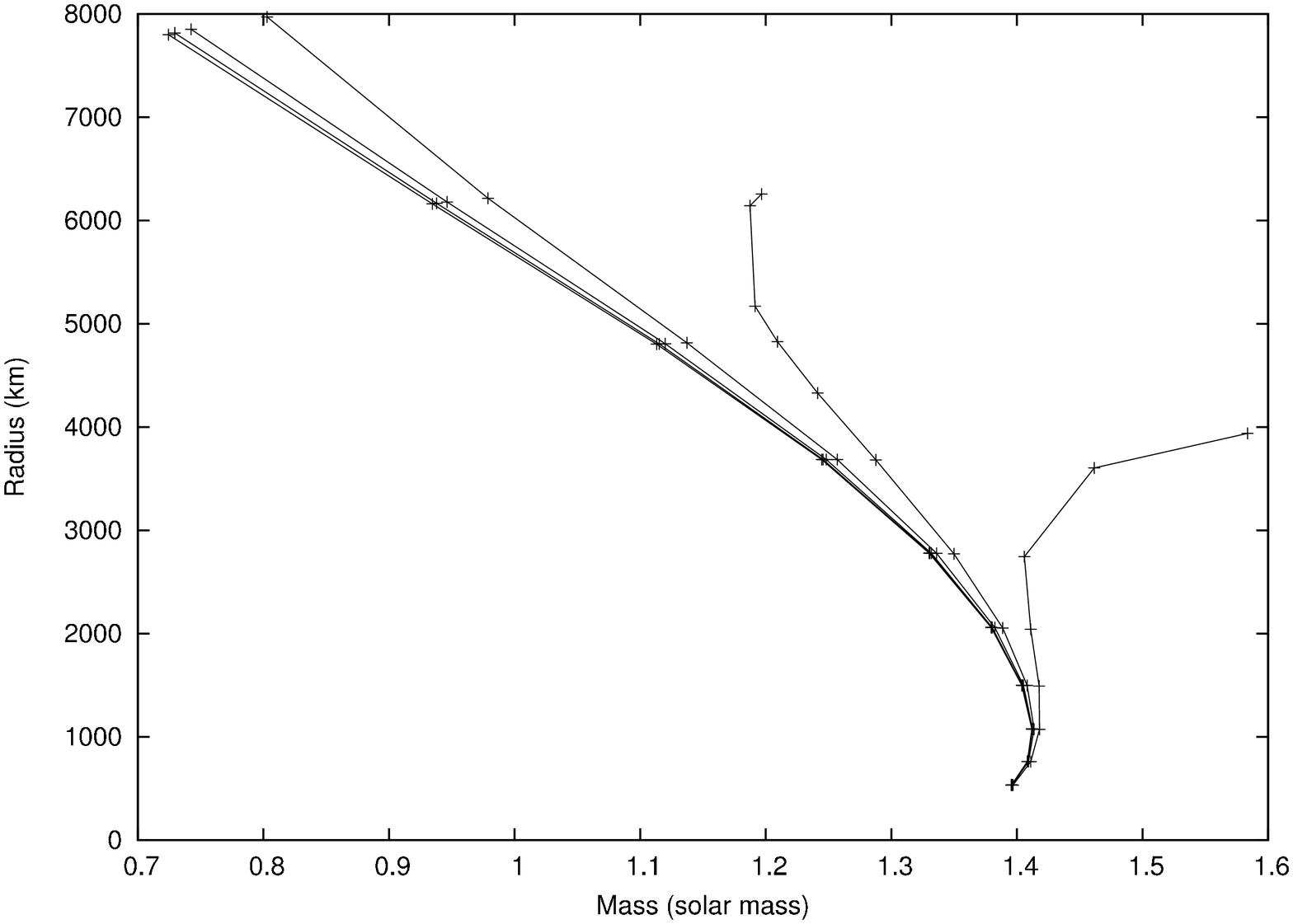}
        \caption{
Mass-radius relation of white dwarfs with spin period $1.95$ min and 
different magnetic fields.
For the lines from left (bottom) to right (top) the corresponding surface fields range
$0-0$, $(3.6-45)\times 10^{10}$, $(7.2-82)\times 10^{10}$, $(1.8-14.8)\times 10^{11}$, 
$(8.5-26.3)\times 10^{11}$ and $(2.4-4.7)\times 10^{12}$ G respectively for the respective range of central density. 
The corresponding central fields range $0-0$, $(3.3-148.3)\times 10^{11}$, $(6.1-270.4)\times 10^{11}$,
$(1.23-48.9)\times 10^{12}$, $(4.5-86.4)\times 10^{12}$ and $(1.9-15.5)\times 10^{13}$ G respectively. Here the radius is basically the equatorial radius.
See text for other details.
}
        \label{massrad}
         \end{figure}

Figure \ref{massrad} describes the $M-R$ relation obtained by LORENE keeping the field restriction in AR Sco 
in mind. We also choose the star to be rotating with period $1.95$ min, as is for
the white dwarf in AR Sco. 
In order to obtain a solution, hence $M$ and $R$, by LORENE, first we assign a 
central density ($\rho_c$) 
and a certain parameterisation of the magnetic fields (and hence current), namely ``charge function". 
Each $M-R$ curve's the sequence of top-most to bottom-most points follow the sequence of increasing
$\rho_c$ for a given charge function. 
Now with the change of charge function, we obtain the family of $M-R$ relation, shown in Fig. \ref{massrad}.
For a given charge function, increasing $\rho_c$ corresponds to increasing $B_s$ and $B_c$ for
self-consistency\footnote{However, 
the trend of increasing fields with
the increase of $\rho_c$ would change for the very high field stars.}, as given in Fig. \ref{massrad} caption.
The readers interested to understand more details are referred to the LORENE manual.
As seen in Fig. \ref{massrad}, generally $M$ decreases as $\rho_c$ decreases along with
increasing $R$ for low (or non) magnetic white dwarfs, which bring them in the very similar trend, except 
for the top two curves.
The maximum $\rho_c$, for completeness, is chosen to be $2\times 10^{11}$ g cm$^{-3}$ which leads to a unstable 
zone in the $M-R$ curve where $M$ further decreases with increasing $\rho_c$. Upto 
$\rho_c\sim 2\times 10^{10}$ g cm$^{-3}$, however, the $M-R$ curve remains stable. The chosen minimum 
$\rho_c$ is $6.3\times 10^6$ g cm$^{-3}$, except for top two curves where they are 
$2\times 10^7$ g cm$^{-3}$ (top but one) and $1.3\times 10^8$ g cm$^{-3}$ (top-most).
For $B_s>2\times 10^{12}$, i.e. for top two curves, 
the trend changes. This is because at lower densities,
gravitational power decreases rendering a bigger size of white dwarfs which however would be massive
due to the additional effects of stronger magnetic pressure unlike the nonmagnetic (low magnetic) cases.
For $B_s<10^{11}$ G and $B_c<10^{12}$ G, $M-R$ relations are practically same as of non-magnetic white dwarfs. 

However, Fig. \ref{massrad} shows that in the mass range $0.8-1.29M_\odot$, the increase in radius due 
to such stronger fields could be at most
a factor of 1.6 only which would render about a factor of 2.6 increase only in $L_{\dot{\nu}}$.
Hence, apparently the fields in AR Sco seem not to be playing any practical role in determining its 
mass and radius currently, which eventually controls its observed luminosity. Nevertheless, 
note importantly that all the inferences are based on LORENE with purely poloidal field geometry,
which need not be the general configuration.

\section{Possible evolution of AR Sco}
\label{evo}

The present high values of magnetic field and spin rate of AR Sco 
could be either due to their high natal values or due to an increase in 
these values during its binary evolution. Since, as a class, white dwarfs
are not known to have high spin and magnetic fields at their birth  (unlike
neutron stars), most likely, AR Sco acquired its present values of
magnetic field and spin during its binary evolution. 
The class of objects with white dwarf as a primary and a late
type low mass star as a secondary is  known as cataclysmic variables (CV)
(see \citealt{warner}, for a review) and, in \citealt{dmr},
we have postulated the concept of a white dwarf increasing its mass and
magnetic field by accretion,  eventually leading to a 
B-WD. There are, indeed,  evidences to suggest that white dwarfs 
in CV configuration have significantly higher magnetic fields
than detached binaries (\citealt{lbert}) and that white
dwarfs in CVs are more massive than the expected canonical 
values (\citealt{zoro}). The single degenerate scenario
of the production of type Ia supernova postulates the rapid
accretion of matter onto a white dwarf (\citealt{wi}).

The detailed investigation of the process of accretion and the 
increase of mass in CVs, however, faces several difficulties 
like accretion leading to nova eruptions (thus decreasing the white dwarf
mass) and the accumulated shells erupting and getting ejected.
Detailed calculations show that thermal-timescale mass transfers
are not effective in increasing the masses of white dwarfs in CVs
(see, e.g., \citealt{ll}). The discovery of a fast rotating
and high magnetic field white dwarf in AR Sco opens up the
possibility of episodic increase of mass in a CV and in the
following we sketch a tentative scheme of repeated episodes of
mass accretion as a reason for the high magnetic field and spin rate
in AR Sco. This mechanism can quite possibly lead to a B-WD.
An accretion driven scenario for the spin-up of AR Sco 
has been considered by \cite{bes}
and here we present a comprehensive picture of the increase of mass,
spin, and magnetic filed in white dwarfs, which might eventually 
lead to a B-WD. It is important to note the 
observational evidence for transitions between rotation/spin-powered and
accretion-powered phases in a binary millisecond pulsar (\citealt{papi})
--- further strongly motivating us to explore similar possibility in
the case of a white dwarf pulsar.

\begin{figure}
        \includegraphics[scale=0.5,angle=0]{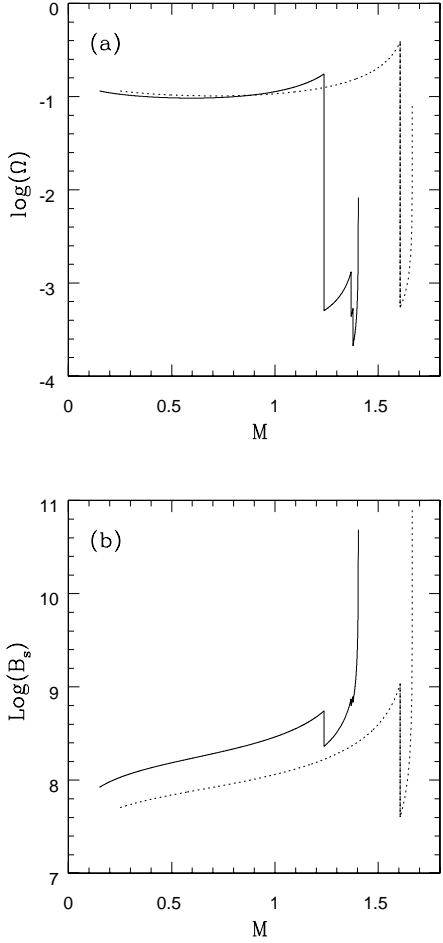}
        \caption{Time evolution of (a) angular velocity in sec$^{-1}$, (b) magnetic field
in G, as functions of mass in units of solar mass. The solid curves correspond
to the case with $n=3$, $m=2.7$, $\rho=0.05$ gm cm$^{-3}$, $l=1.5$ and dotted
curves correspond to the case with $n=3$, $m=2$, $\rho=0.1$ gm cm$^{-3}$, $l=2.5$.
Other parameters are $k=10^{-14}$ CGS, $\dot{M}=10^{-8}M_\odot$Yr$^{-1}$, 
$\alpha=10$ degree and $R=10^4$ km at $t=0$.
}
        \label{timev}
\end{figure}

We assume that AR Sco actually was a binary which accretes mass from its companion.
 As a result of mass gain, its gravitational power might have increased rendering 
decreasing radius and from the flux freezing increasing of its any initial magnetic   
fields. On the other hand, 
due to the conservation of angular momentum, its angular velocity and hence spin frequency
varies with the change of mass and radius, whether decreasing or increasing, that
depends on the trend of moment of inertia (i.e. $MR^2$), as depicted in Fig. \ref{timev}
(will be discussed below).
In addition, at some point,
accretion might also have been inhibited temporarily due to appreciable increase in 
fields creating significant outward force to oppose infall (this also could be 
due to its entering in the propeller phase, see, e.g., \citealt{ghosh}). As a result, 
it would behave as a spin-powered pulsar (we interchangeably use the phrase
rotation-powered and spin-powered) with increasing spin period $P_s$ and with time $P_s$ 
turns out to be what we see today.

\begin{figure}
        \includegraphics[scale=0.5,angle=0]{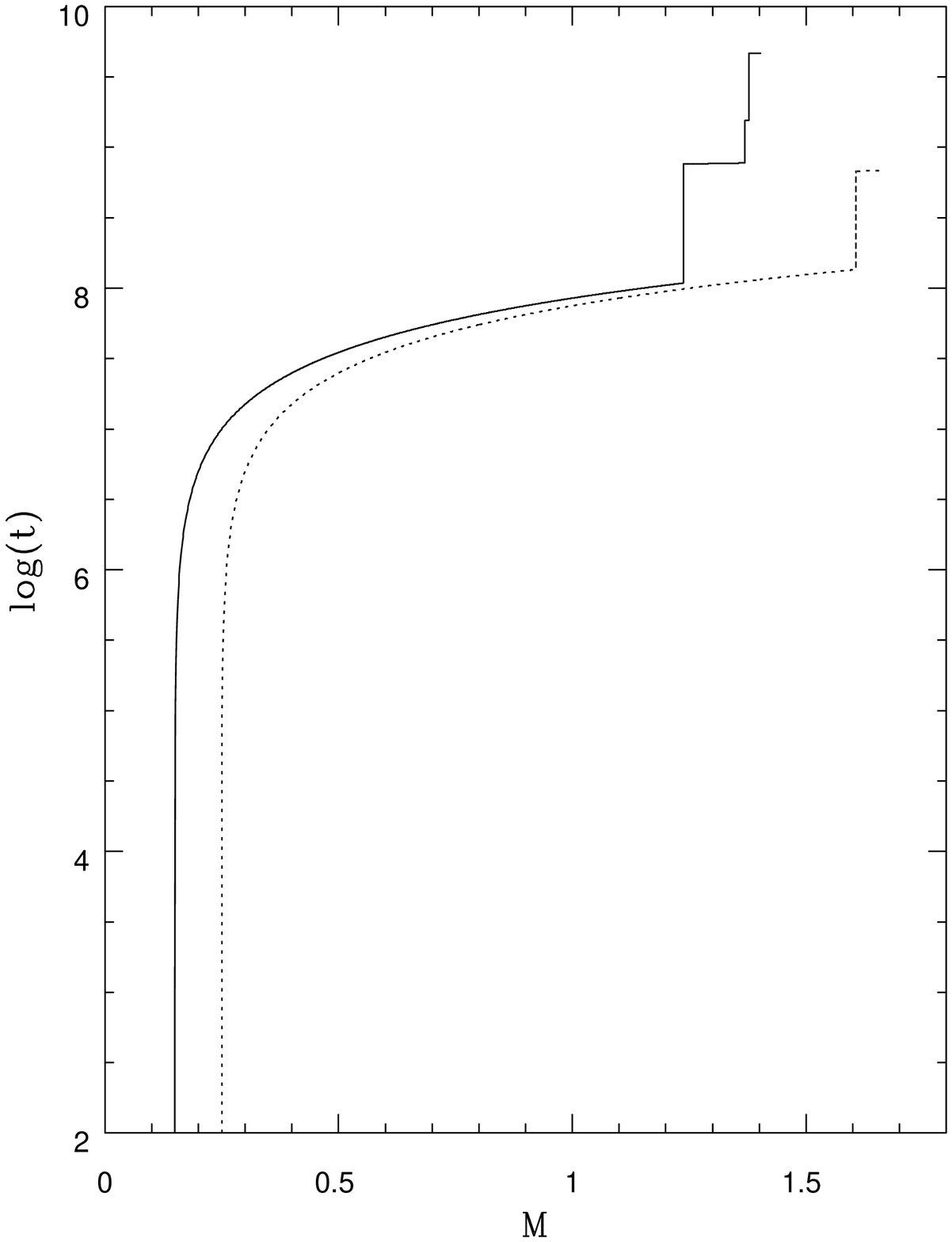}
        \caption{
Time taken in Yr to evolve the mass and magnetic fields of white dwarfs 
shown in Fig. \ref{timev}.
}
        \label{time}
\end{figure}

After $\gtrsim 10^8$ Yr (as verified in \S3, Fig. \ref{time}), due to continuous radio emission, its angular velocity as well as fields 
would decay significantly (as is the case for a rotating dipole, see, e.g., \citealt{jack}) 
and eventually radiation would stop. However, at this stage it would start 
accreting again. Moreover, because of the shrinkage in the binary system resulting from
gravitational wave emission, the white dwarf by this time would acquire 
stronger gravitational power to accrete matter from the companion more efficiently.
A simple estimate (\citealt{bers}) argues that the time $\Delta t$ taken for the decrease of separation between
white dwarf of mass $M_1$ and its companion of mass $M_2$ from $r_1$ to $r_2$ is given by
\begin{equation}
\Delta t=\frac{5}{256}\frac{c^5}{G^3}\left(\frac{r_1^4-r_2^4}{M_1 M_2(M_1+M_2)}\right),
\end{equation}
where $G$ is Newton's gravitation constant.
Therefore, for $M_1=4\times 10^{33}$ gm, $M_2=2\times 10^{33}$ gm, $r_1=5\times 10^5$ km and $r_2=10^5$ km, 
$\Delta t$ turns out to be below $10^8$ Yr
and hence significant binary shrinkage is justified during the repeated episodes of mass accretion
and spin-powered phases.
However, such a white dwarf (might turn out to be a B-WD 
candidate) would also radiate
gravitational wave due to its non-spherical shape as a consequence of magnetic effects
and spinning nature (as will be discussed more quantitatively in \S 4). 

Hence, after restarting accretion, the whole cycle described above would repeat again and again 
rendering much stronger fields eventually. Note that once the binary shrinkage takes
place significantly,
the decay phase of angular velocity and magnetic fields gets abolished and 
frequency and fields both would start increasing uninterruptedly, until the companion is exhausted.
As a consequence, the underlying white dwarf
would not follow the theory of non-magnetic white dwarfs and would cross the
Chandrasekhar's limit. Eventually, it would deviate from Chandrasekhar's $M-R$ trajectory to 
B-WD's trajectory, as demonstrated earlier by \cite{dmr}. 

Based on a toy model, the above speculative proposition can be examined. There are two phases:
accretion-powered and rotation-powered.
 There are three conservation laws controlling the accretion-powered phase:
linear and angular momenta conservation and conservation of 
magnetic flux, around the stellar surface, which could be closer to the inner edge of accretion disc
depending on the field strength, given by 
\begin{eqnarray}
\nonumber
l\Omega(t)^2 R(t) &=& \frac{GM(t)}{R(t)^2},\\
\nonumber
I(t)\Omega(t)&=&{\rm constant},\\
B_s(t)R(t)^2&=&{\rm constant},
\label{conv}
\end{eqnarray}
where $l$ takes care of inequality due to dominance of gravitational force
over the centrifugal force in general (so that terms with pressure and magnetic fields 
are parameterized in $l$; larger the value of $l$, stronger the effects of pressure and fields
over rotation assumed), $I$ is the moment of inertia of star and
$\Omega$ is the angular velocity of the star which includes the additional contribution
acquired due to accretion as well. Solving equations in (\ref{conv}) simultaneously, we obtain 
the time evolution of radius (or mass), magnetic field and angular velocity during accretion.
Accretion stops when 
\begin{eqnarray} 
-\frac{GM}{R^2}=\frac{1}{\rho}\frac{d}{dr}\left(\frac{B^2}{8\pi}\right)|_{r=R}
\sim -\frac{B_s^2}{8\pi R\rho},
\end{eqnarray} 
where $\rho$ is the density of inner edge of disc.

If the magnetic field is of dipole nature, $\dot{\Omega}\propto\Omega^3$ for a
fixed magnetic field (see, e.g., \citealt{bmrao}), where over-dot implies time 
derivative. Generalizing it, for the present
purpose we assume $\dot{\Omega}=k\Omega^n$ with $k$ being constant. Therefore,
during the phase of spin-powered pulsar (when accretion inhibits, even 
temporarily), without having explicit knowledge of field geometry,
time evolution of angular velocity and surface magnetic field are given by
\begin{eqnarray}
\Omega=\left[\Omega_0^{1-n}-k(1-n)(t-t_0)\right]^{\frac{1}{1-n}},
\label{omd}
\end{eqnarray}
\begin{eqnarray}
B_s=\sqrt{\frac{5c^3Ik\Omega^{n-m}}{R^6\sin^2\alpha}},
\label{bd}
\end{eqnarray}
where $\Omega_0$ is the angular velocity at the beginning of spin-powered phase
(when accretion just stops) at time $t=t_0$.
$k$ is fixed in order to constrain $B_s$ at $t=t_0$, at the beginning of first spin-powered phase, 
which is known from the evolution of fields in the preceding accretion-powered phase.
Note that $n=m=3$ corresponds to dipole field, hence $m$ represents the deviation from dipolar field 
particularly for $n=3$.

Figure \ref{timev} shows a couple of sample possible evolutions of angular
velocity and magnetic field with mass. Note that the mass of star is
varying with time (shown in Fig. \ref{time} below) and hence considered in the horizontal axis to describe 
the evolution. $\dot{M}$ in all the accretion-powered phases is chosen to be $10^{-8}M_\odot$Yr$^{-1}$, 
which is slightly higher than that of a typical
intermediate polar, but an order of magnitude lower $\dot{M}$ would also suffice our purpose.
Other parameters, mentioned in the Fig. \ref{timev} caption, are some of their typical representative values.
In both the cases, initial larger $\Omega$ 
with accretion is seen to be dropped significantly during spin-powered phase (when accretion stops and
hence no change of mass),
followed by a phase of its increasing trend. The case with solid line shows a few such
phases/cycles with dip: they are determined by the choice of $n$ and $m$. Similar trend is seen in surface magnetic field profiles
with a sharp increasing trend (with value $\sim 10^{11}$ G) at the last 
cycle. This corresponds to the increase of $B_c$ as well leading to a B-WD. At the end of evolution,
it could be left out as a super-Chandrasekhar white dwarf and/or a SGR/AXP candidate with a higher spin
frequency.
Of course, they are just representative samples and they may
depend on many other factors and hence they apparently do not match 
exactly with what is expected to happen in AR Sco itself.

\begin{figure}
        \includegraphics[scale=0.32,angle=270]{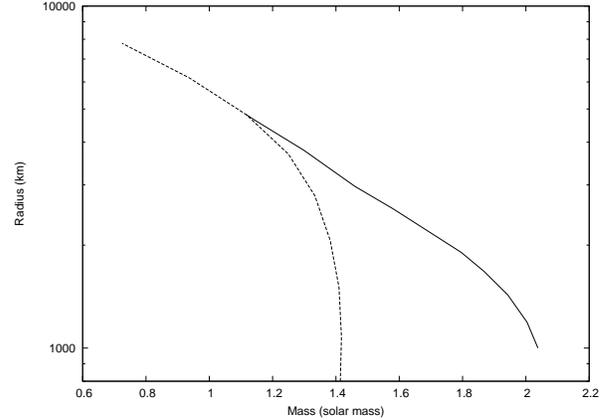}
        \caption{
Mass-radius relation of Chandrasekhar's non-magnetised (or weakly magnetised)
white dwarfs (dotted line) and that of evolutionary track of accreting 
highly magnetised (poloidal) B-WDs (solid line). Here the radius for B-WDs is basically 
the equatorial radius.
}
        \label{trajsft}
\end{figure}

Figure \ref{trajsft} shows how the $M-R$ trajectory with the increase of fields 
could deviate from Chandrasekhar's to B-WD's ending with an eventual larger
limiting mass. This is similar to what was argued earlier (\citealt{dmr}) in
the presence of even stronger fields. With the increase of mass due to accretion,
radius decreases which leads to the increase of magnetic fields assuming conservation
of magnetic flux. Increasing magnetic fields however creates increasing outward pressure
which is able to oppose stronger gravitational field even at lower density.
This eventually leads to massive super-Chandrasekhar white dwarfs even below 
$\rho_c=2\times 10^{10}$ gm cm$^{-3}$. Figure \ref{time} confirms that the above mentioned increase
of mass $\Delta M$, with $\dot{M}\sim 10^{-8}M_\odot~{\rm Yr}^{-1}$, would
complete in $\sim 10^9$ Yr, which is quite legitimate given
the current age of Universe. It is clear by comparing Fig. \ref{time} with Fig. \ref{timev}
that the dip of $\Omega$ and $B_s$ in the latter corresponds to the sharp
increase of $t$ in the former, as in the spin-powered phase there is no change of mass.

As indicated in Fig. \ref{timev} qualitatively, repeating the cycles described above 
will reveal lower and lower $P_s$ apart from higher and higher fields,
ending up forming a fast spinning B-WD, which might behave as a SGR/AXP. 
For quantitative estimates, one should explore more rigorous model.
Indeed, several
SGRs/AXPs have been argued to be fast spinning B-WDs and those sources are 
successfully explained without invoking extraordinarily high unobserved yet magnetic fields (\citealt{bmrao}).
As B-WDs are expected to be about 100 pc away, many AR Sco like objects are expected to be
seen in future astronomical missions like Square Kilometer Array (SKA).

\section{Emission of gravitational wave and star-quake}
\label{grav}

Neutron stars are already proposed to be the candidates of continuous gravitational wave signal due to
their quadrupolar nature.
Such a signal is possible to emit by a tri-axial compact star rotating around a principle 
axis of inertia due to its quadrupole moment characterized by the amplitude (\citealt{palomba})
\begin{equation}
h_+(t)=h_0\left(\frac{1+\cos^2\alpha_0}{2}\right)\cos\Phi(t),\,\,h_{\times}(t)=h_0\cos\alpha_0\sin\Phi(t),
\label{h+x}
\end{equation}
where $\alpha_0$ is the inclination of the star's rotation axis with respect to the line of sight and
\begin{equation}
h_0=\frac{4\pi^2G}{c^4}\frac{I_{zz}\epsilon}{P_s^2 d},
\label{ho}
\end{equation}
$\Phi(t)$ is the signal phase function,
$I_{zz}$ is the moment of inertial about z-axis, $\epsilon$ is the measure of ellipticity of the star and
$d$ is the distance of the star from the 
detector.

As rotating B-WDs are ellipsoid and could rotate faster than their standard counter-parts,
they also could be plausible candidates for continuous gravitational wave signal
(see also \citealt{heyl,schram2}). A B-WD with mass $\sim 2M_\odot$, 
polar radius $\sim 700$ km, $P_s\sim 1$ sec (\citealt{sathya}), $\epsilon\sim 5\times 10^{-4}$ and  
at $\sim 100$ pc away from us 
would produce $h_0\sim 10^{-22}$, which is within the sensitivity 
of the Einstein@Home search for early Laser Interferometer Gravitational Wave Observatory 
(LIGO) S5 data (\citealt{palomba}). 
However, a firm confirmation of gravitational wave emission can be provided
by detectors more sensitive in their frequency range like 
Deci-hertz Interferometer Gravitational wave Observatory or 
Big Bang Observer (DECIGO/BBO) (\citealt{yagi}).
In fact, if the B-WD's polar radius is $\sim 2000$ km with $P_s\sim 10$ sec and other parameters intact,
even DECIGO/BBO can detect it with $h_0\sim 10^{-23}$. 
Nevertheless, high magnetic field rotating white dwarfs approaching B-WDs would be common and
it is possible that such white dwarfs of radius $\sim 7000$ km, $P_s\sim 20$ sec and $d\sim 10$ pc 
will have a $h_0 \gtrsim 10^{-22}$ which is detectable by Laser Interferometer Space 
Antenna (LISA) (\citealt{moore}). Note that the
chosen value for $\epsilon$ needs to be realised based on rigorous theory.

Also if observed spin-down is totally due to gravitational wave emission, the absolute upper limit 
of signal is (\citealt{palomba})
\begin{equation}
h_0^{sd}=8.06\times 10^{-19} I_{45}d_{\rm kpc}^{-1}\sqrt{\frac{|\dot{\nu_s}/{\rm Hz~sec^{-1}}|}{\nu_s/{\rm Hz}}},
\label{hsd}
\end{equation}
where $\nu_s=1/P_s$, $I_{45}$ is the moment of inertia in units of $10^{45}$ gm cm$^2$ and 
$d_{\rm kpc}$ is the distance of
the source in units of kpc. For the above mentioned B-WD with $P_s=1$ sec but of size $1000$ km,
$h_0^{sd}\sim 10^{-20}$ for 
$\dot{P_s}\sim 10^{-14}$ sec~sec$^{-1}$ (similar but slightly lower compared to that of AR Sco). 

After significant accretion, 
the gravitational power may dominate the centrifugal effects of the core significantly to pull 
it to a less oblate shape due to significant decrease in radius, 
thereby stressing it.
The release of such stresses would lead to
a sudden decrease in the moment of inertia and, correspondingly, by the conservation of angular momentum
it would suddenly rotate fast such that
\begin{equation}
\frac{\Delta I}{I}=-\frac{\Delta \Omega}{\Omega},
\end{equation}
see \citealt{ostg,pac}.
This will lead to a star quake in B-WDs with the increase of gravitational energy 
given by
\begin{equation}
\Delta E_{G}\sim\frac{GM^2}{R}\frac{\Delta R}{R}.
\end{equation}
Here $\Delta$ denotes the change of respective quantities.
The corresponding gain of rotational energy is
\begin{equation}
\Delta E_{rot}=-4\pi^2 I\frac{\Delta P_s}{P_s^3}.
\end{equation}
For a $2M_\odot$ B-WD with $R\sim 1000$ km, $\Delta E_{G}\sim 5\times 10^{51}\Delta P_s/P_s$
and $\Delta E_{rot}\sim -1.6\times 10^{51}\Delta P_s/P_s^3$, hence rotational energy could be 
explained by available gravitational energy, particularly for $P_s\sim 1$ sec, 
and observed flares/gaint-flares/outbursts in SGRs/AXPs
of energy $10^{43}-10^{46}$ ergs could be explained as star quakes. 

\section{Conclusion}

The idea of B-WD is in the literature for quite sometime (see also 
\citealt{ost}), however without any direct proof of their existence.
The major motivation of introducing B-WD by us was to explain peculiar
over-luminous type Ia supernovae. Later on, other sources like SGRs/AXPs,
a white dwarf pulsar were also explained under the B-WD premise. 
Nevertheless, there is no direct observational evidence for them yet.
Of course, due to the small  size, they were speculated to be of low
luminosity (\citealt{dm14}). Also, as total gravitational force of B-WDs
does not appear to change significantly (for similar radius, there
may be about $50\%$ increase of mass), in the presence of 
high magnetic field, thermal pressure and hence luminosity may be decreased
(which however needs to be checked in a rigorous calculation).

We have explored the possibility of AR Sco being the seed of B-WD. 
Although observed data of AR Sco currently at hand do not seem to violate 
Chandrasekhar's theory, we have shown that weak magnetic fields
in the underlying white dwarf may enhance during accretion, which
may deviate it from Chandrasekhar's $M-R$ trajectory and leading
to a spinning B-WD in the life-time of Universe. 
This is however an exploratory study based on a simple model, which
should be re-investigated in detail in future.

We have also touched upon the issue of gravitational wave possibly emitted
from B-WDs. While LISA appears to be very appropriate to detect them,
even LIGO S5 may do so. All in all, the present work is the first attempt
to unfold the issue of direct observational evidences of B-WDs. 
Although not evident yet, it indicates a plausible path and future prospect of
their direct detection.

\section{Acknowledgment}

We thank Varun Bhalerao, A. Gopakumar and Nirupam Roy for discussion. 
The work was partly supported by the project with research Grant No. 
ISTC/PPH/BMP/0362.


\begin{thebibliography}{68}
\expandafter\ifx\csname natexlab\endcsname\relax\def\natexlab#1{#1}\fi

\bibitem[{{Allen} {et~al}\mbox{.}(2006){Allen}, {Dunn}, {Fabian}, {Taylor}, \&
  {Reynolds}}]{Allen2006}
{Allen} S.~W., {Dunn} R.~J.~H., {Fabian} A.~C., {Taylor} G.~B., {Reynolds}
  C.~S., 2006, MNRAS, 372, 21

\bibitem[{{Asai} {et~al}\mbox{.}(1998){Asai}, {Dotani}, {Hoshi}, {Tanaka},
  {Robinson}, \& {Terada}}]{Asai1998}
{Asai} K., {Dotani} T., {Hoshi} R., {Tanaka} Y., {Robinson} C.~R., {Terada} K.,
  1998, PASJ, 50, 611

\bibitem[{{Baganoff} {et~al}\mbox{.}(2003){Baganoff}, {Maeda}, {Morris},
  {Bautz}, {Brandt}, {Cui}, {Doty}, {Feigelson}, {Garmire}, {Pravdo}, {Ricker},
  \& {Townsley}}]{Baganoff2003}
{Baganoff} F.~K. {et~al.}, 2003, ApJ, 591, 891

\bibitem[{{Balbus} \& {Hawley}(1998)}]{Balbus1998}
{Balbus} S.~A., {Hawley} J.~F., 1998, Reviews of Modern Physics, 70, 1

\bibitem[{{Belloni} {et~al}\mbox{.}(2000){Belloni}, {Klein-Wolt}, {M{\'e}ndez},
  {van der Klis}, \& {van Paradijs}}]{Belloni2000}
{Belloni} T., {Klein-Wolt} M., {M{\'e}ndez} M., {van der Klis} M., {van
  Paradijs} J., 2000, \aap, 355, 271

\bibitem[{{Bildsten} \& {Rutledge}(2000)}]{Bildsten2000}
{Bildsten} L., {Rutledge} R.~E., 2000, \apj, 541, 908

\bibitem[{{Bisnovatyi-Kogan}, {Zel'dovich} \&
  {Nadezhin}(1972){Bisnovatyi-Kogan}, {Zel'dovich}, \&
  {Nadezhin}}]{Bisnovatyi-Kogan1972}
{Bisnovatyi-Kogan} G.~S., {Zel'dovich} Y.~B., {Nadezhin} D.~K., 1972, Soviet
  Astronomy, 16, 393

\bibitem[{{Blandford} \& {Begelman}(1999)}]{Blandford1999}
{Blandford} R.~D., {Begelman} M.~C., 1999, MNRAS, 303, L1

\bibitem[{{Blondin}, {Mezzacappa} \& {DeMarino}(2003){Blondin}, {Mezzacappa},
  \& {DeMarino}}]{Blondin2003}
{Blondin} J.~M., {Mezzacappa} A., {DeMarino} C., 2003, \apj, 584, 971

\bibitem[{{Bondi}(1952)}]{Bondi1952}
{Bondi} H., 1952, MNRAS, 112, 195

\bibitem[{{Cackett} {et~al}\mbox{.}(2010){Cackett}, {Brown}, {Cumming},
  {Degenaar}, {Miller}, \& {Wijnands}}]{Cackett2010}
{Cackett} E.~M., {Brown} E.~F., {Cumming} A., {Degenaar} N., {Miller} J.~M.,
  {Wijnands} R., 2010, \apjl, 722, L137

\bibitem[{{Campana} {et~al}\mbox{.}(1998){Campana}, {Colpi}, {Mereghetti},
  {Stella}, \& {Tavani}}]{Campana1998}
{Campana} S., {Colpi} M., {Mereghetti} S., {Stella} L., {Tavani} M., 1998,
  \aapr, 8, 279

\bibitem[{{Chakrabarti}(1989)}]{chakrabarti1989}
{Chakrabarti} S.~K., 1989, \apj, 347, 365

\bibitem[{{Chakrabarti}(1999)}]{Chakrabarti1999}
{Chakrabarti} S.~K., 1999, \aap, 351, 185

\bibitem[{{Chen} {et~al}\mbox{.}(1998){Chen}, {Cui}, {Frank}, {King}, {Livio},
  \& {Zhang}}]{Chen1998}
{Chen} W., {Cui} W., {Frank} J., {King} A., {Livio} M., {Zhang} S., 1998, in
  American Institute of Physics Conference Series, Vol. 431, American Institute
  of Physics Conference Series, {Holt} S.~S., {Kallman} T.~R., eds., pp.
  347--350

\bibitem[{{Chevalier} \& {Imamura}(1982)}]{Chevalier1982}
{Chevalier} R.~A., {Imamura} J.~N., 1982, \apj, 261, 543

\bibitem[{{D'Angelo} {et~al}\mbox{.}(2015){D'Angelo}, {Fridriksson},
  {Messenger}, \& {Patruno}}]{DAngelo2015}
{D'Angelo} C.~R., {Fridriksson} J.~K., {Messenger} C., {Patruno} A., 2015,
  \mnras, 449, 2803

\bibitem[{{Das} \& {Sharma}(2013)}]{Das2013}
{Das} U., {Sharma} P., 2013, MNRAS, 435, 2431

\bibitem[{{Foglizzo}, {Galletti} \& {Ruffert}(2005){Foglizzo}, {Galletti}, \&
  {Ruffert}}]{Foglizzo2005}
{Foglizzo} T., {Galletti} P., {Ruffert} M., 2005, \aap, 435, 397

\bibitem[{{Foglizzo} {et~al}\mbox{.}(2007){Foglizzo}, {Galletti}, {Scheck}, \&
  {Janka}}]{Foglizzo2007}
{Foglizzo} T., {Galletti} P., {Scheck} L., {Janka} H.-T., 2007, \apj, 654, 1006

\bibitem[{{Foglizzo} {et~al}\mbox{.}(2012){Foglizzo}, {Masset}, {Guilet}, \&
  {Durand}}]{Foglizzo2012}
{Foglizzo} T., {Masset} F., {Guilet} J., {Durand} G., 2012, Phys. Rev. Lett.,
  108, 1103

\bibitem[{{Foglizzo} \& {Tagger}(2000)}]{Foglizzo2000}
{Foglizzo} T., {Tagger} M., 2000, A\&A, 363, 174

\bibitem[{{Frank}, {King} \& {Raine}(2002){Frank}, {King}, \&
  {Raine}}]{Frank2002}
{Frank} J., {King} A., {Raine} D.~J., 2002, {Accretion Power in Astrophysics:
  Third Edition}

\bibitem[{{Garcia} {et~al}\mbox{.}(2001){Garcia}, {McClintock}, {Narayan},
  {Callanan}, {Barret}, \& {Murray}}]{Garcia2001}
{Garcia} M.~R., {McClintock} J.~E., {Narayan} R., {Callanan} P., {Barret} D.,
  {Murray} S.~S., 2001, ApJ, 553, L47

\bibitem[{{Gu} \& {Foglizzo}(2003)}]{gu2003}
{Gu} W.-M., {Foglizzo} T., 2003, \aap, 409, 1

\bibitem[{{Guilet} \& {Foglizzo}(2012)}]{Guilet2012}
{Guilet} J., {Foglizzo} T., 2012, MNRAS, 421, 546

\bibitem[{{Hanke} {et~al}\mbox{.}(2012){Hanke}, {Marek}, {M{\"u}ller}, \&
  {Janka}}]{Hanke2012}
{Hanke} F., {Marek} A., {M{\"u}ller} B., {Janka} H.-T., 2012, \apj, 755, 138

\bibitem[{{Herant} {et~al}\mbox{.}(1994){Herant}, {Benz}, {Hix}, {Fryer}, \&
  {Colgate}}]{Herant1994}
{Herant} M., {Benz} W., {Hix} W.~R., {Fryer} C.~L., {Colgate} S.~A., 1994,
  \apj, 435, 339

\bibitem[{{Holzer} \& {Axford}(1970)}]{Holzer1970}
{Holzer} T.~E., {Axford} W.~I., 1970, \araa, 8, 31

\bibitem[{{Igumenshchev} \& {Narayan}(2002)}]{Igumenshchev2002}
{Igumenshchev} I.~V., {Narayan} R., 2002, \apj, 566, 137

\bibitem[{{Illarionov} \& {Sunyaev}(1975)}]{Illarionov1975}
{Illarionov} A.~F., {Sunyaev} R.~A., 1975, \aap, 39, 185

\bibitem[{{Iwakami} {et~al}\mbox{.}(2009){Iwakami}, {Kotake}, {Ohnishi},
  {Yamada}, \& {Sawada}}]{Iwakami2009}
{Iwakami} W., {Kotake} K., {Ohnishi} N., {Yamada} S., {Sawada} K., 2009, \apj,
  700, 232

\bibitem[{{Langer}, {Chanmugam} \& {Shaviv}(1981){Langer}, {Chanmugam}, \&
  {Shaviv}}]{Langer1981}
{Langer} S.~H., {Chanmugam} G., {Shaviv} G., 1981, \apjl, 245, L23

\bibitem[{{Loewenstein} {et~al}\mbox{.}(2001){Loewenstein}, {Mushotzky},
  {Angelini}, {Arnaud}, \& {Quataert}}]{Loewenstein2001}
{Loewenstein} M., {Mushotzky} R.~F., {Angelini} L., {Arnaud} K.~A., {Quataert}
  E., 2001, ApJ, 555, L21

\bibitem[{{McCrea}(1956)}]{McCrea1956}
{McCrea} W.~H., 1956, \apj, 124, 461

\bibitem[{{Menou} {et~al}\mbox{.}(1999){Menou}, {Esin}, {Narayan}, {Garcia},
  {Lasota}, \& {McClintock}}]{Menou1999}
{Menou} K., {Esin} A.~A., {Narayan} R., {Garcia} M.~R., {Lasota} J.-P.,
  {McClintock} J.~E., 1999, \apj, 520, 276

\bibitem[{{Mignone} {et~al}\mbox{.}(2007){Mignone}, {Bodo}, {Massaglia},
  {Matsakos}, {Tesileanu}, {Zanni}, \& {Ferrari}}]{Mignone2007}
{Mignone} A., {Bodo} G., {Massaglia} S., {Matsakos} T., {Tesileanu} O., {Zanni}
  C., {Ferrari} A., 2007, ApJS, 170, 228

\bibitem[{{Molteni}, {Sponholz} \& {Chakrabarti}(1996){Molteni}, {Sponholz}, \&
  {Chakrabarti}}]{Molteni1996}
{Molteni} D., {Sponholz} H., {Chakrabarti} S.~K., 1996, \apj, 457, 805

\bibitem[{{Molteni}, {T{\'o}th} \& {Kuznetsov}(1999){Molteni}, {T{\'o}th}, \&
  {Kuznetsov}}]{mtk1999}
{Molteni} D., {T{\'o}th} G., {Kuznetsov} O.~A., 1999, \apj, 516, 411

\bibitem[{{Mukhopadhyay}(2002)}]{Mukhopadhyay2002}
{Mukhopadhyay} B., 2002, International Journal of Modern Physics D, 11, 1305

\bibitem[{{Mukhopadhyay}(2009)}]{Mukhopadhyay2009}
{Mukhopadhyay} B., 2009, \apj, 694, 387

\bibitem[{{Mukhopadhyay} {et~al}\mbox{.}(2003){Mukhopadhyay}, {Ray}, {Dey}, \&
  {Dey}}]{Mukhopadhyay2003}
{Mukhopadhyay} B., {Ray} S., {Dey} J., {Dey} M., 2003, \apjl, 584, L83

\bibitem[{{Nakayama}(1992)}]{nakayama1992}
{Nakayama} K., 1992, \mnras, 259, 259

\bibitem[{{Narayan}, {Garcia} \& {McClintock}(1997){Narayan}, {Garcia}, \&
  {McClintock}}]{Narayan1997}
{Narayan} R., {Garcia} M.~R., {McClintock} J.~E., 1997, ApJ, 478, L79

\bibitem[{{Narayan} \& {Yi}(1995)}]{Narayan1995}
{Narayan} R., {Yi} I., 1995, \apj, 452, 710

\bibitem[{{Nobuta} \& {Hanawa}(1994)}]{nobuta1994}
{Nobuta} K., {Hanawa} T., 1994, \pasj, 46, 257

\bibitem[{{Paczy{\'n}sky} \& {Wiita}(1980)}]{Paczynsky1980}
{Paczy{\'n}sky} B., {Wiita} P.~J., 1980, A\&A, 88, 23

\bibitem[{{Proga} \& {Begelman}(2003)}]{Proga2003}
{Proga} D., {Begelman} M.~C., 2003, \apj, 582, 69

\bibitem[{{Psaltis}, {Belloni} \& {van der Klis}(1999){Psaltis}, {Belloni}, \&
  {van der Klis}}]{Psaltis1999}
{Psaltis} D., {Belloni} T., {van der Klis} M., 1999, \apj, 520, 262

\bibitem[{{Remillard} \& {McClintock}(2006)}]{Remillard2006}
{Remillard} R.~A., {McClintock} J.~E., 2006, \araa, 44, 49

\bibitem[{{Reynolds} \& {Miller}(2009)}]{Reynolds2009}
{Reynolds} C.~S., {Miller} M.~C., 2009, \apj, 692, 869

\bibitem[{{Ruffert}(1994{\natexlab{a}})}]{Ruffert1994a}
{Ruffert} M., 1994{\natexlab{a}}, \apj, 427, 342

\bibitem[{{Ruffert}(1994{\natexlab{b}})}]{Ruffert1994c}
{Ruffert} M., 1994{\natexlab{b}}, Astronomy \& Astrophysics Supplement, 106

\bibitem[{{Ruffert} \& {Arnett}(1994)}]{Ruffert1994b}
{Ruffert} M., {Arnett} D., 1994, \apj, 427, 351

\bibitem[{{Rybicki} \& {Lightman}(1986)}]{Rybicki1986}
{Rybicki} G.~B., {Lightman} A.~P., 1986, {Radiative Processes in Astrophysics}.
  p. 400

\bibitem[{{Ryu}, {Chakrabarti} \& {Molteni}(1997){Ryu}, {Chakrabarti}, \&
  {Molteni}}]{Ryu1997}
{Ryu} D., {Chakrabarti} S.~K., {Molteni} D., 1997, \apj, 474, 378

\bibitem[{{Sakashita}(1974)}]{sakashita1974a}
{Sakashita} S., 1974, Astrophysics and Space Science, 26, 183

\bibitem[{{Sakashita} \& {Yokosawa}(1974)}]{sakashita1974b}
{Sakashita} S., {Yokosawa} M., 1974, Astrophysics and Space Science, 31, 251

\bibitem[{{Saxton}(2002)}]{Saxton2002}
{Saxton} C.~J., 2002, \pasa, 19, 282

\bibitem[{{Scheck} {et~al}\mbox{.}(2008){Scheck}, {Janka}, {Foglizzo}, \&
  {Kifonidis}}]{Scheck2008}
{Scheck} L., {Janka} H.-T., {Foglizzo} T., {Kifonidis} K., 2008, \aap, 477, 931

\bibitem[{{Sedov}(1946)}]{Sedov1946}
{Sedov} L.~I., 1946, Journal of Applied Mathematics and Mechanics, 10, 241

\bibitem[{{Sharma} {et~al}\mbox{.}(2007){Sharma}, {Quataert}, {Hammett}, \&
  {Stone}}]{Sharma2007}
{Sharma} P., {Quataert} E., {Hammett} G.~W., {Stone} J.~M., 2007, \apj, 667,
  714

\bibitem[{{Sharma}, {Quataert} \& {Stone}(2008){Sharma}, {Quataert}, \&
  {Stone}}]{Sharma2008}
{Sharma} P., {Quataert} E., {Stone} J.~M., 2008, MNRAS, 389, 1815

\bibitem[{{Shvartsman}(1971)}]{Shvartsman1971}
{Shvartsman} V.~F., 1971, Soviet Astronomy, 15, 377

\bibitem[{{Stone}, {Pringle} \& {Begelman}(1999){Stone}, {Pringle}, \&
  {Begelman}}]{Stone1999}
{Stone} J.~M., {Pringle} J.~E., {Begelman} M.~C., 1999, \mnras, 310, 1002

\bibitem[{{Taylor}(1950)}]{Taylor1950}
{Taylor} G., 1950, Royal Society of London Proceedings Series A, 201, 159

\bibitem[{{van der Klis}(2004)}]{vanderklis2004}
{van der Klis} M., 2004, ArXiv Astrophysics e-prints

\bibitem[{{Yuan} \& {Narayan}(2014)}]{Yuan2014}
{Yuan} F., {Narayan} R., 2014, \araa, 52, 529

\end{thebibliography}


\begin{thebibliography}{30}
\expandafter\ifx\csname
natexlab\endcsname\relax\def\natexlab#1{#1}\fi

\bibitem[Belyaev et al.(2015)]{belya} Belyaev V. B., Ricci P., Simkovic F., Adam J., Tater M., Truhlik E.,
2015, Nuc. Phys. A, 937, 17
\bibitem[Beskrovnaya \& Ikhsanov(2016)]{bes} Beskrovnaya N. G., Ikhsanov N. R., 2016, Proceeding of the ``Stars from collapse to collapse", 
Nizhnij Arkhyz, Karachai-Cherkessian Republic, Special Astrophysical
Observatory of the Russian Acad. of Sci.,
October 3-7, 2016; arXiv:1612.07831v1
\bibitem[Bera \& Bhattacharya(2016)]{bb15} Bera P., Bhattacharya D., 2016, MNRAS, 456, 3375
\bibitem[Bertschinger \& Taylor(2015)]{bers} Bertschinger E., Taylor E. F., 2015, in AW Physics Macros,
GravWaves150909v1, 1, 15
\bibitem[Bocquet et al.(1995)]{lorene1} Bocquet M., Bonazzola S., Gourgoulhon E., Novak J., 1995,
A\&A, 301, 757
\bibitem[Bonazzola et al.(1993)]{lorene2} Bonazzola S., Gourgoulhon E., Salgado M., Marck J. A.,
1993, A\&A, 278, 421
\bibitem[Chandrasekhar(1935)]{chandra35} Chandrasekhar S., 1935, MNRAS, 95, 207
\bibitem[Das \& Mukhopadhyay(2013)]{prl} Das U., Mukhopadhyay B., 2013, Phys. Rev. Lett., 110, 071102
\bibitem[Das \& Mukhopadhyay(2014)]{dm14} Das U., Mukhopadhyay B., 2014, MPLA, 
29, 1450035
\bibitem[Das \& Mukhopadhyay(2015)]{dm15} Das U., Mukhopadhyay B., 2015, JCAP, 05, 016
\bibitem[Das, Mukhopadhyay \& Rao(2013)]{dmr} Das U., Mukhopadhyay B., Rao A. R., 2013, ApJ, 767, L14
\bibitem[Franzon \& Schramm(2015)]{schram} Franzon B., Schramm S., 2015, Phys. Rev. D, 92, 083006
\bibitem[Franzon \& Schramm(2017)]{schram2} Franzon B., Schramm S., 2017, 
MNRAS, 467, 4484
\bibitem[Ghosh(1995)]{ghosh} Ghosh P., 1995, ApJ, 453, 411
\bibitem[Liebert et al.(2015)]{lbert} Liebert J., Ferrario L., Wickramasinghe 
D.~T., Smith P.~S., 2015, ApJ, 804, 93
\bibitem[Jackson(1999)]{jack} Jackson J. D., 1999, in Classical Electrodynamics, 3rd Edn., 
John Wiley \& Sons
\bibitem[Heyl(2000)]{heyl} Heyl J. S., 2000, MNRAS, 317, 310
\bibitem[Liu \& Li(2016)]{ll} Liu W.-M., Li, X.-D., 2016, ApJ, 832, 80
\bibitem[Liu, Zhang \& Wen(2014)]{liu} Liu H., Zhang X., Wen, D., 2014, Phys. Rev. D, 89, 104043
\bibitem[Marsh et al.(2016)]{marshnature} Marsh T. R., et al., 2016, Nature, 537, 374
\bibitem[Mereghetti(2012)]{mereghetti} Mereghetti S., 2012, in Proceedings of the 26th Texas 
Symposium on Relativistic Astrophysics, Sao Paulo, December 16-20, 2012; arXiv1304.4825
\bibitem[Moore, Cole \& Berry(2015)]{moore} Moore C. J., Cole R. H., \& Berry C. P. L., 2015, Class. Quant. Grav.,
32, 015014
\bibitem[Mukhopadhyay \& Rao(2016)]{bmrao} Mukhopadhyay B., Rao A. R., 2016, JCAP, 05, 007
\bibitem[Ostriker \& Gunn(1969)]{ostg} Ostriker J. P., Gunn J. E., 1969, ApJ, 157, 1395
\bibitem[Ostriker \& Hartwick(1968)]{ost} Ostriker J. P., Hartwick F. D. A., 
1968, ApJ, 153, 797
\bibitem[Paczynski(1990)]{pac} Paczynski B., 1990, ApJ, 365, L9
\bibitem[Palomba et al.(2013)]{palomba} Palomba C., 2012, the LIGO Scientific Collaboration and the 
Virgo Collaboration -- Proceedings of the Recontres de Moriond, 2011; arXiv:1201.3176
\bibitem[Papitto et al.(2013)]{papi} Papitto A., et al., 2013, Nature, 501, 517
\bibitem[Subramanian \& Mukhopadhyay(2015)]{sathya} Subramanian S., Mukhopadhyay B., 2015, MNRAS, 454, 752
\bibitem[Tong \& Xu(2012)]{tongxu12} Tong H., Xu R. X., 2012, ApJ, 757, L10
\bibitem[Warner(1995)]{warner} Warner B., 1995, Cam. Astrophys. Ser., 28
\bibitem[Whelan \& Iben (1973)]{wi} Whelan J, Iben I.~Jr., 1973, ApJ, 186, 1007
\bibitem[Yagi \& Seto (2011)]{yagi} Yagi K., Seto N. 2011, Phys. Rev. D, 83(4), 044011
\bibitem[Zhang \& Harding(2000)]{bing} Zhang B., Harding A. K., 2000, 535, L51
\bibitem[Zorotovic, Schreiber \& G{\"a}nsicke(2011)]{zoro} 
Zorotovic M., Schreiber M.~R., \& G{\"a}nsicke B.~T., 2011, A\&A, 536, A42 








\end{thebibliography}
\end{document}